\documentclass[twocolumn,prl,amsmath,amssymb,unsortedaddress,superscriptaddress,showpacs]{revtex4}
\usepackage{latexsym,epsfig,graphicx}
\usepackage{dcolumn}
\usepackage{bm}
\usepackage{color} 
\usepackage{graphicx}
\usepackage{subfigure}
\usepackage{tikz}
\usetikzlibrary{matrix}

\begin{document}

\title[Note on the "size" of Schroedinger cats] {Note on the "size" of Schroedinger cats\footnote{This note was originally written (but not widely circulated) in April 2009,in response to a preprint version of ref. \cite{JK1}.This slightly expanded  version is posted now because of its relevance to ref. \cite{G}.} }


\author{Anthony J. Leggett}
\affiliation{Department of Physics, University of Illinois at Urbana-Champaign, Urbana, IL 61801, USA}


\maketitle


 In ref.~\cite{JK1} Korsbakken et al.define a number which they 
 propose as a measure of the degree of "macro/mesoscopic distinctness" of 
 two states which may be involved in a quantum superposition. Crudely
 speaking,this number (which they call $\Delta N_{tot}$; for brevity I will refer 
 to it as W) is the minimum number of single particles which one needs to 
 shift to orthogonal states in order to go from one state to the other.More 
 technically, in a basis k in which the single-particle occupation numbers 
 $n_k$ are diagonal, we have
 \begin{eqnarray}
   W=\sum_k \left| \Delta n_k\right|
 \end{eqnarray}  
 
where $\Delta n_k$ is the difference in $\left\langle n_k\right\rangle $ in the two states compared.
They then evaluate W for those experiments on flux qubits in the existing 
literature which have been interpreted as evidence for "macroscopic quantum
coherence", and find that the maximum value which has been obtained to date
is only of order 5000. They comment that "[this number is] well short of
anything that could reasonably be considered meso-or macroscopic".

While I agree with Korsbakken et al.'s evaluation of W for the flux 
qubit experiments, the purpose of this note is to point out that for at 
least one type of situation where I think most people would agree that the
two states in question are "macro-(not meso-!)scopically distinct" the 
value of $W$ is actually $\_$smaller$\_$ than in those experiments. However, 
examination of the the example used suggests strongly that the use 
of the value of W as a measure of "macroscopic distinctness" is rather
unnatural,and suggests a different figure of merit, which I evaluate for 
the flux qubit case.

Let us provisionally define two different states of a physical system
as "macroscopically distinct" if they can be resolved by the unaided human
senses over a "reasonable" period,say for definiteness one second;and 
further more define two states as "mesoscopically distinct" if they can be 
similarly resolved under some magnification $X^{-1}$. (The reader is invited to 
pause at this point and decide,before reading further,whether he/she 
agrees with the first definition, and to choose a value of $X$ in the second
definition). Now, the minimum spatial difference which can be resolved by 
the unaided human eye is about 1.5 microns~\cite{X}. Let us err on the 
side of conservatism and thus consider a particle, say for definiteness of 
LiF, of diameter 5 microns,traversing its own diameter over a period of a
second; then according to the above definition this state of the particle
is macroscopically distinct from its rest state.(Similarly, the two 
corresponding states of a particle of diameter $5X$ microns are
"mesoscopically" distinct.)

Now let us make an approximate evaluation of the number W for this 
pair of states.It is convenient to introduce quite generically (also for the flux-
qubit case) a characteristic velocity $v_0$ such that the contribution of 
each kind of particle to W can be written in the form $N(v/v_0)$ where N is 
the total number of such particle involved. Thus, for the flux-qubit case,
 which involves only electrons, N is of order $10^9$-$10^{10}$ and $v_0$ is,apart 
from a factor of order unity,the electron Fermi velocity \cite{JK1}.

The LiF particle considered contains about $8*10^{14}$ electrons and 
about 2.2 times that number of nucleons. Consider first the electrons. Since
in the rest state they occupy only Bloch-wave states in filled bands, and 
the only difference in the moving state is that the bands are shifted, it
is clear that the value of $v_0$ is simply of order $h/ma$, where a is the linear
dimension of the cubic unit cell; thus $v_0$ is about $3*10^8$ cm/sec (which is 
comparable, not surprisingly, to the Fermi velocity of a typical metal).

Before assessing the effect of the nucleons, let us consider one 
possible objection to the above line of argument: It may be argued that
the only way in which we can in fact tell the two states of the particle 
apart is because we actually discriminate not the two velocities but the 
two positions occupied after one second, for which W (or even the 
contribution to it from the electrons) is definitely of order N. Even
supposing that this argument is psychophysically correct, the problem is
that in view of the different nature of the geometry there is no 
corresponding number one can define for the flux qubit, so the point 
seems moot.

In considering the nucleons, it is at first sight tempting to take $v_0$ 
to be of the order of the root-mean-square atomic velocity, which is about 
three orders of magnitude smaller than the electron Fermi velocity; this
would give a contribution to W several orders of magnitude greater than 
that of the electrons. {However-and this} is the crucial point-if we take the 
definition given by Korsbakken et al. seriously, we should actually take $v_0$ 
to be of the order of the rms velocity of the individual  {$\_$nucleons$\_$} within 
the nucleus! Since the latter is about one order of magnitude larger than 
the electron Fermi velocity, and the number of nucleons is only about 
twice that of the electrons, the nucleon contribution to W is only about 
0.2 of the electronic one.

Putting these results together, we see that the value of W for the two 
macroscopically distinct states of our LiF particle is approximately 
$1.2*8*10^{14}*(5*10^{-4}/3*10^8)$, i.e.around 1600-$\_$smaller$\_$  than that for
"best" flux qubit! It is amusing that by reducing the size to the "minimum
visible" value of 1.5 microns we can actually get a value of W which is 
smaller than 3.

Of course,the reader's first and entirely natural reaction to this 
surprising result will no doubt be that I have "cheated" by not treating 
the motion of the individual nucleons in terms of that of the composite
nuclei. But my point is that the "cheating" is also implicit in the 
calculation carried out in {ref. \cite{JK1}:} If one wishes, perhaps reasonably, to 
argue that when a particular type of particle is bound into a composite 
then the quantity W should be evaluated in terms of the $\Delta n_k$ of the 
composite, then one must be prepared to treat the flux-qubit case in terms 
of the  $\_$center-of-mass$\_$  coordinates of the Cooper pairs, which are, after 
all, perfectly good "composites". (One can bolster this point by a 
thought-experiment in which we sweep the system across the "BCS-BEC 
crossover"; at the BEC end it is clear that it is the COM degree of freedom 
of the $\_$molecules$\_$ which is relevant, and current theory and experiment 
suggests no discontinuity in the crossover).

Before proceeding, I should note that in p.3,col.2.para.2 of their extended
discussion in ref.\cite{JK2} Korsbakken et al. also briefly refer to the Cooper pairs 
and conclude that "looking at Cooper-pair modes rather than single-electron 
modes  does not change the effective cat size". However, the quantity they 
consider (the $\Delta N_{k,-k}$ of their eqn.(6)) bears no relation to anything that 
could reasonably be called the COM wave function of the pairs, and would 
give a very small cat size even in the BEC limit; I therefore believe that it is 
irrelevant to the present discussion.

Suppose,first,that the external flux is half a flux quantum ("symmetric" 
case), so that the (classically degenerate) many-body states which are 
superposed correspond to center-of-mass (kinematic)  angular momentum 
$1/2*\hbar$ and $-1/2*\hbar$ respectively and the corresponding COM wave 
functions are mutually orthogonal. Then if one takes the reasonable 
point of view that the fraction of all the electrons which are "bound into 
Cooper pairs" is $N_p/N$ where $N_p\approx$$N*(\Delta/\epsilon_F)$ is the single macroscopic 
eignvalue of the reduced 2-particle density matrix, then the resultant value
of W (call it $W_{CP}$) is of order $10^6$-$10^7$. Since the corresponding number 
for the LiF particle (i.e.that obtained by treating the basic constituents as 
electrons and nuclei) is about $10^5$, we see that the states superposed in 
the flux qubit should still count as macroscopically distinct.

The "asymmetric" case, which is realized in the experiment reported in 
ref. \cite{G}, is trickier, since while the  $\_$many-body$\_$ wave functions of the two 
states  whose superposition is confirmed \footnote{In the sense that a macrorealistic account is excluded.}  are of course mutually 
orthogonal to a very high degree of accuracy, the corresponding Cooper-pair 
COM wave functions are not. In this case the most natural extension of the 
definition of $W_{CP}$ would seem to be simply $N_p$ times $(1-K)$ where $K$ is 
the overlap of the two COM wave functions (this would seem to correspond 
most closely to the definition of W for the  $\_$single-particle$\_$ case in ref. \cite{JK1}). 
However, it is clear that a more quantitative discussion is desirable. 

   I am grateful to Birgitta Whaley for sending me a copy of ref.[1] before 
   publication, to Frances Wang for information on human visual acuity and to
   Markus Aspelmeyer for helpful discussions. I also thank Mao-chuang Yeh 
   for pointing out a numerical error in the original version, and for useful 
   comments.

\end{document}